\documentstyle[psfig,epsfig,12pt]{article}

\newcommand{\be}{\begin{equation}}
\newcommand{\ee}{\end{equation}}
\newcommand{\bea}{\begin{eqnarray}}
\newcommand{\eea}{\end{eqnarray}}

\title{Quantum Chaos at Finite Temperature}

\author{
L.A.~Caron$^{a}$, H.~Jirari$^{a}$, 
H.~Kr\"{o}ger$^{a}\footnote{Corresponding author, 
Email: hkroger@phy.ulaval.ca}$,
X.Q.~Luo$^{b,c}$, \\
G.~Melkonyan$^{a}$ 
and K.J.M.~Moriarty$^{d}$  
\\ [2mm]
{\small\sl $^{a}$D\'{e}partement de Physique, Universit\'{e} Laval, Qu\'{e}bec, Qu\'{e}bec G1K 7P4, Canada} \\ 
{\small\sl $^{b}$CCAST (World Laboratory), P.O. Box 8730, Bejing 100080, China} \\ 
{\small\sl $^{c}$Departement of Physics, Zhongshan University, Guangzhou 510275, China} \\
{\small\sl $^{d}$Department of Mathematics, Statistics and Computer Science,} \\
{\small\sl Dalhousie University, Halifax, N.S. B3H 3J5, Canada} \\ 
}

\begin{document} 
\maketitle

\noindent Abstract: \\
We use the quantum action to study quantum chaos at finite temperature. We present a numerical study of a classically chaotic 2-D Hamiltonian system 
- harmonic oscillators with anharmonic coupling. We construct the quantum action non-perturbatively and find temperature dependent quantum corrections in the action parameters. We compare Poincar\'e sections of the quantum action at finite temperature with those of the classical action.

\setcounter{page}{0}

\newpage

\section{Introduction} 
Deterministic chaos has been observed in a huge number of 
phenomena in macroscopic i.e. classical physics. But chaotic phenomena 
were also found in microscopic physics ruled by quantum mechanics -- an area presently under very active investigation. For reviews see Refs.\cite{Gutzwiller,Nakamura,Blumel,Stockmann,Haake}.
For example, the hydrogen atom in a strong magnetic field 
shows strong irregularities in its spectrum \cite{Friedrich}.
Irregular patterns have been found in the wave functions of the quantum mechanical stadium billard \cite{McDonald}.
Billard like boundary conditions have been realized experimentally 
in mesoscopic quantum systems, like quantum dots and quantum corrals, formed by atoms in semi-conductors \cite{Stockmann}. 

Classical chaos theory can not simply be taken over to quantum physics (due to Heisenberg's uncertainty relation). Hence workers have tried to characterize quantum chaos in ways alternative to classical chaos. 
A fruitful approach is Gutzwiller's periodic orbit theory \cite{Gutzwiller}
which predicts level densities in the semi-classical regime (Rydberg states in atoms). Wintgen \cite{Wintgen} applied it to the diamagnetic hydrogen system and was able to extract periodic orbit information from experimental level densities. Another successful route is the analysis of level densities in the context of random matrix theory. A conjecture by Bohigas et al. \cite{Bohigas} states that the signature of a classical chaotic system is described by random matrix theory and its spectral density follows a Wigner distribution.

Strictly speaking, there is no universally accepted definition of quantum chaos. Many roads have been travelled searching for "irregular" signals in quantum systems. In the quest for quantum chaos, it has been popular to examine systems 
which exhibit classical chaos.
However, Jona-Lasinio et al. \cite{Jona:92} found chaotic behavior in a quantum system without classical counterpart, i.e., a quantum many-body system undergoing multiple resonant tunneling.
Jona-Lasinio and Presilla \cite{Jona:96} further studied 
quantum many-body systems of interacting bosons in the thermodynamic limit. 
They found chaotic signals when the interaction strength drives the system away from integrability.

Signals of quantum chaos have been investigated also in fermionic many-body systems, in particular by looking at the Poisson versus Wigner level spacing distributions \cite{Jacquod:97,Georgeot:97}.
Georgeot and Shepelyansky \cite{Georgeot:98,Georgeot:99,Shepelyansky:00} 
studied a spinglass model of spin $1/2$ particles and looked at the transition from the integrable to the chaotic regime, when tuning a random transverse magnetic field. 
Casetti et al. \cite{Casetti:98} considered the large $N$ limit of $N$-component $\phi^{4}$ oscillators in the presence of an external field. They compare classical chaos with quantum chaos, computed in mean field theory. They observe a strong suppression of quantum chaos due to quantum corrections moving the system away from a hyperbolic fixed point responsable for classical chaos.
Habib et al. \cite{Habib:98} have studied in chaotic systems the relation between classical and quantum expectation values. They found divergence between the two after a time scale $t_{\hbar} \sim ln(C/\hbar)$. Decoherence reduced this discrepancy.   

However, the study of quantum chaos lacks from the following shortcomings:
(i) In quantum chaos, we have no "local" information of the degree of chaoticity, being available in classical chaos via Lyapunov exponents and Poincar\'e sections from phase space.
(ii) Also little is known about the role of temperature in quantum chaos.
For example, the analysis of level densities is insensitive to temperature.
To overcome those problems,
we adopt the strategy of building a bridge between quantum physics and classical physics. This means a relation involving the quantum transition amplitude and the classical action. Starting from the path integral, the following "bridges" have been considered:

(i) {\it Sum over classical paths}. The quantum mechanical transition amplitude can be expressed via the 
path integral. In certain cases this path integral can be expressed as a sum over classical paths only.
This holds, e.g. for the harmonic oscillator. 
Unfortunately, such relation holds only in a few exceptional cases \cite{Schulman}.

(ii) {\it Gutzwiller's trace formula}. 
Gutzwiller \cite{Gutzwiller} has established a relation between the 
density of states of the quantum system and a sum over 
classical periodic orbits (periodic orbit quantisation). The trace formula has been applied in the semi-classical regime (e.g. highly excited states of atom). Wintgen \cite{Wintgen} applied it to the diamagnetic hydrogen system and extracted periodic orbit information from experimental level densities.

(iii) {\it Effective action}. 
Another possibility is to use the convential effective action \cite{Jona:64,Coleman:73}. An effective action has been also considered at finite temperature \cite{Dolan}. The effective action has been introduced in such a way that it gives an expectation value $<\phi>=\phi_{class}$ which corresponds to the classical trajectory and which minimizes the potential energy (effective potential). 
Thus one can obtain the ground state energy of the quantum system from its effective potential. Because the effective action has a mathematical structure similar to the classical action, it looks like the ideal way to bridge the gap from quantum to classical physics and eventually solve the quantum chaos and quantum instanton problem.
Cametti et al.\cite{Cametti:99} have 
computed in Q.M. the effective action perturbatively via loop ($\hbar$) expansion and studied quantum chaos in the 2-D anharmonic oscillator. 
Perturbation theory forces to terminate such series at a finite (low) order.
The difficulty lies in the estimation of the remainder. 
This poses a problem in the context of chaos, because 
chaotic systems are known to exhibit sensitivity not only to initial conditions but also to parameters of the system. 
There are higher loop corrections to the effective potential $V^{eff}$ as well as to the mass renormalisation $Z$. 
The kinetic term of the effective action is given by an asymptotic series of higher time derivatives (infinite series of increasing order). 
The problem in interpreting $\Gamma$ as effective action is that
those higher time derivatives require more intial/boundary conditions than the classical action.

(iv) {\it Bohm's interpretation of quantum mechanics}.
An interesting approach has been taken by Iacomelli and Pettini \cite{Iacomelli:96}.
They have used Bohm's \cite{Bohm:52} interpretation of quantum mechanics, which expresses the Schr\"{o}dinger equation in polar form. The radial part satisfies a continuity equation of a velocity field and the phase satisfies a Hamilton-Jacobi equation. The "trajectories" of the quantum system are equivalent to Lagrangian trajectories of the fluid velocity field.
This allows to apply the concept of classical chaos theory and in particular compute Lyapunov exponents. In Ref.\cite{Iacomelli:96} this framework has been used to study the chaotic behavior of the hydrogen atom in an external magnetic field and different regimes of quantum chaoticity have been observed.

\section{Quantum Action}
\noindent In the following we will explore the quantum action \cite{Jirari:01a} as bridge between classical and quantum physics. 
This action has a mathematical structure like the classical action, and takes into account quantum effects (quantum fluctuations) via renormalized action parameters. The quantum action has the following attractive properties \cite{Jirari:01b}:
(i) It can be computed non-perturbatively, via numerical simulations. 
(ii) It is defined for arbitrary finite temperature.
(iii) In the zero-temperature limit 
analytic expressions exist for the quantum action, provided that the classical potential is sufficiently symmetric (parity symmetric in D=1 and rotationally symmetric in D=2,3). 
(iv) In a numerical study of the 1-D anharmonic oscillator, 
we searched for the presence of higher order time derivative terms in the kinetic term of the quantum action. No signal for such term was found \cite{Jirari:01b}. 
The quantum action has proven useful for rigorous definition and 
quantitative computation of quantum instantons \cite{Jirari:01c}. 
Let us recall the definition of the quantum action, as proposed 
in Ref.\cite{Jirari:01a} \\ 
\noindent {\it Conjecture}:
For a given classical action $S = \int dt \frac{m}{2} \dot{x}^{2} - V(x)$ 
there is a quantum action 
$\tilde{S} = \int dt \frac{\tilde{m}}{2} \dot{x}^{2} - \tilde{V}(x)$, 
which allows to express the transition amplitude by
\begin{equation}
\label{DefQuantumAction}
G(x_{fi},t_{fi}; x_{in},t_{in}) = \tilde{Z} 
\exp [ \frac{i}{\hbar} \left. \tilde{S}[\tilde{x}_{cl}] 
\right|_{x_{in},t_{in}}^{x_{fi},t_{fi}} ] .
\end{equation}
Here $\tilde{x}_ {cl}$ denotes the classical path, such that the action $\tilde{S}(\tilde{x}_{cl})$ 
is minimal (we exclude the occurrence of conjugate points or caustics). 
$\tilde{Z}$ denotes the normalisation factor corresponding to $\tilde{S}$. 
Eq.(\ref{DefQuantumAction}) is valid with 
the {\em same} action $\tilde{S}$ for all sets of 
boundary positions $x_{fi}$, $x_{in}$ for a given time interval $T=t_{fi}-t_{in}$. 
The parameters of the quantum action depend on the time $T$.  
Any dependence on $x_{fi}, x_{in}$ enters via the trajectory 
$\tilde{x}_ {cl}$. $\tilde{Z}$ depends on the action parameters and $T$, 
but not on $x_{fi}, x_{in}$.

The quantum action at finite time $T$ can be interpreted as action at 
finite temperature.
The partition function in quantum mechanics requires to go over from real time to imaginary time and to impose periodic boundary conditions \cite{Kapusta:89}.
In Ref.\cite{Jirari:01b} we have shown that the expectation value of a quantum mechanical observable $O$ at thermodymical equilibrium can be expressed in terms of the quantum action along its classical trajectory from $x_{in}$, $\beta_{in}=0$ to $x$, $\beta$, given by
\bea
&& \tilde{\Sigma}_{\beta} 
\equiv 
\tilde{S}_{\beta}[\tilde{x}_{cl}]|_{x_{in},0}^{x,\beta} = 
\int_{0}^{\beta} d \beta' ~ \frac{m}{2 \hbar^{2}} (\frac{d \tilde{x}_{cl}}{d \beta'})^{2} 
+ \tilde{V}(\tilde{x}_{cl})
\nonumber \\
&& \frac{\delta \tilde{S}_{\beta}}{\delta x}[\tilde{x}_{cl}] = 0 
\nonumber \\
&& \tilde{x}_{cl}(\beta_{in})=x_{in}, ~ \tilde{x}_{cl}(\beta) = x ~ ,
\eea
where $T$ denotes the absolute value of imaginary time. The parameters $\beta$, temperature $\tau$ and $T$ are related by 
$\beta = \frac{1}{k_{B} {\tau}} = T/\hbar$.
In numerical studies with the quartic potential in 1-D we found the parameters of the quantum action to be temperature dependent \cite{Jirari:01a,Jirari:01c,Jirari:01b}.

\section{Analytic form of quantum action at zero temperature}
\label{sec:AnalyticForm}
Why are we interested in the zero-temperature limit?
First, from the experimental point of view, the ground state properties of a quantum system are often much better known than excited states. 
Second, zero temperature corresponds to large time and 
from the point of view of chaos theory, 
it is interesting to study dynamical behavior at large times. Examples:
(a) Behavior of particles moving in storage rings, (b) In classical Hamiltonian chaos, there are quasi-periodic trajectories, remaining for a long time in the vicinity of regular islands. What is the corresponding behavior in quantum mechanics?
(c) What is the quantum analogue of the KAM theorem?
(d) The construction of Poincar\'e sections and Lyapunov exponents involves 
long time trajectories.
Likewise the quantum analogue of Poincar\'e sections and Lyapunov 
exponents will require long time evolutions.

In the following we work in imaginary time. In order to avoid notational confusion, we continue to use the time parameter $T$.
In particular, $T \to 0$ denotes the high temperature limit, where the quantum action approaches its classical counter part. $T \to \infty$ denotes the zero-temperature limit, where the quantum action is determined by the Feynman-Kac formula.  
In Ref.\cite{Jirari:01b} the following analytic transformation law at temperature zero has been derived in 1-D relating the classical action $S=\int dt \frac{1}{2} m \dot{x}^{2} + V(x)$ to the quantum action $\tilde{S} = \int dt \frac{1}{2} \tilde{m} \dot{x}^{2} + \tilde{V}(x)$,
\be
\label{TransformFeynKac}
2 m(V(x) - E_{gr}) =  
2 \tilde{m}(\tilde{V}(x) - \tilde{v}_{0}) 
- \frac{\hbar}{2} \frac{ \frac{d}{dx} 2 \tilde{m} (\tilde{V}(x) - \tilde{v}_{0})}
{ \sqrt{2 \tilde{m}( \tilde{V}(x) - \tilde{v}_{0} ) } } \mbox{sgn}(x) ~ .
\ee
Moreover, the following relation between the ground state wave function and the quantum action holds
\be
\label{GroundStateFeynKac}
\psi_{gr}(x) = \frac{1}{N} ~ e^{ - \int_{0}^{|x|} dx' ~ 
\sqrt{2 \tilde{m}( \tilde{V}(x') - \tilde{v}_{0} ) }/\hbar }, ~
E_{gr} = \tilde{v}_{0} ,
\ee
Eqs.(\ref{TransformFeynKac},\ref{GroundStateFeynKac}) are valid under the assumption  of a parity symmetric potential $\tilde{V}(x)$ with a unique minimum at $x=0$, $\tilde{V}(0)=\tilde{v}_{0}$.
Similar laws hold for rotationally symmetric potentials in D=2 and 3.
For a number of systems, where the ground state energy and wave function is known analytically, it has been shown \cite{Jirari:01b} that the quantum action reproduces exactly the ground state energy and wave function. 
Examples in 1-D are: Monomial potentials and the inverse square potential. 
Examples for radial motion in 3-D are: Monomial potentials and Coulomb potential
(hydrogen atom) in the sector of non-zero angular momentum.
In particular, the ground state energy coincides with the minimum of the quantum potential $\tilde{V}$ and the location of the minimum of the quantum potential 
coincides with the location of the maximum of the wave function.
One should note that the transformation law, Eq.(\ref{TransformFeynKac}), determines mass $\times$ potential, but not both, mass and potential individually. The determination of the renormalized mass requires an non-perturbative numerical calculation.

In the following we will try to estimate the precision of numerical calculations of the quantum action in the zero-temperature limit (long time limit)
by contrasting it with the analytic expressions of  Eqs.(\ref{TransformFeynKac},\ref{GroundStateFeynKac}).
We consider in 1-D the classical action
\bea
S &=& \int dt \frac{m}{2} \dot{x}^{2} + V(x), ~~~ 
V(x) = v_{2} x^{2} + v_{4} x^{4}, 
\nonumber \\
m &=& 1
\nonumber \\
v_{2} &=& 1 
\nonumber \\
v_{4} &=& 0.01 ~ .
\eea
The Q.M. Hamiltonian has the following ground state energy, 
\be
\label{ExactEnergy}
E_{gr}= 0.710811 ~ .
\ee
For the quantum action we make an ansatz
\be
\tilde{S} = \int dt \frac{\tilde{m}}{2} \dot{x}^{2} + \tilde{V}(x), ~~
\tilde{V}(x) = \tilde{v}_{0} + \tilde{v}_{2} x^{2} + \tilde{v}_{4} x^{4} + \tilde{v}_{6} x^{6} ~ .
\ee
From a fit of the quantum action to Q.M. transition amplitudes 
(for more details see Ref.\cite{Jirari:01a}) we find the action parameters at $T=4.5$
\bea
\label{1DNumData}
&& \tilde{m} = 0.9990(7)
\nonumber \\ 
&& \tilde{v}_{0} = 0.79863(1) 
\nonumber \\
&& \tilde{v}_{2} = 1.013(2) 
\nonumber \\
&& \tilde{v}_{4} = 0.0099(9) 
\nonumber \\
&& \tilde{v}_{6} = 0.0000(3) ~ .
\eea
The parameters $\tilde{m}$, $\tilde{v}_{2}$, $\tilde{v}_{4}$, $\tilde{v}_{6}$ have asymptotically stabilized at about $T=4.5$. 
However, the parameter $\tilde{v}_{0}$ has a slow fall-off behavior for $T \to \infty$. For large $T$ we have fitted $\tilde{v}_{0}$ to the numerical data by                                                          %
\be
\tilde{v}_{0}(T) \leadsto_{T \to \infty} A + \frac{B}{T} + \frac{C}{T^{2}} ~ ,
\ee
and find $A=0.710819$. Thus the asymptotic extrapolation $\tilde{v}_{0}(T \to \infty) = 0.710819$ is very close to $E_{gr}$, given by Eq.(\ref{ExactEnergy}). 
This confirms $E_{gr}=\tilde{v}_{0}$, predicted by Eq.(\ref{GroundStateFeynKac}), i.e. the ground state energy is given by the minimum of the quantum potential.
Let us further check the consistency of other parameters of the quantum potential. This can be done by using 
the transformation law Eq.(\ref{TransformFeynKac}).
Because this transformation law does not determine 
$\tilde{m}$, we have taken $\tilde{m}$ from the numerial data.
The transformation law then determines the parameters of the quantum potential.
We find,
\bea
\tilde{v}_{2} &=& \frac{2}{\tilde{m}} [ m E_{gr}/\hbar]^{2} = 1.01151 
\nonumber \\
\tilde{v}_{4} &=& \frac{2}{3} \frac{\sqrt{2 \tilde{m} \tilde{v}_{2}}}{\hbar} ~ 
[ \tilde{v}_{2} - v_{2} \frac{m}{\tilde{m}} ] = 0.009967 
\nonumber \\
\tilde{v}_{6} &=& \frac{1}{4} \frac{\tilde{v}_{4}^{2}}{\tilde{v}_{2}}
+ \frac{2}{5} \frac{\sqrt{2 \tilde{m} \tilde{v}_{2}}}{\hbar} ~ 
[ \tilde{v}_{4} - v_{4} \frac{m}{\tilde{m}} ]
= 2.89 ~ 10^{-7} ~ .
\eea
This is consistent with the numerical results, given by Eq.(\ref{1DNumData}). 
Finally, let us look at the ground state wave function. 
As reference solution, we have solved the Schr\"{o}dinger equation 
with the original Hamiltonian. Secondly, we computed the wave function 
from the quantum action via Eq.(\ref{GroundStateFeynKac}), using the numerical data given by Eq.(\ref{1DNumData}).
The comparison is given in Fig.[1], showing good agreement. 
In summary of this section, we have shown for the zero-temperature limit
(regime of the Feynman-Kac formula) 
that the transition amplitude can be parametrized by the quantum action.
A comparison of analytical versus numerical computation of the quantum action has shown good agreement.

\section{Quantum chaos in 2-D anharmonic oscillator}
\noindent As is well known 1-D conservative systems with a time-independent Hamiltonian are integrable and do not display classical chaos.
An interesting candidate system to display chaos is the K-system, corresponding to the potential $V=x^{2}y^{2}$. It decribes a 2-D Hamiltonian system, being almost globally chaotic, having small islands of stability \cite{KSystem}. However, according to a suggestion by Pullen and Edmonds \cite{Pullen}, it is more convenient from the numerical point of view to consider the following system, which also exhibits classical chaos. It is a 2-D anharmonic oscillator,
defined  by the following classical action, 
\bea
S &=& \int_{0}^{T} dt ~ \frac{1}{2} m (\dot{x}^{2} + \dot{y}^{2}) 
+ V(x,y), ~~~ V(x,y) = v_{2}(x^{2} + y^{2}) + v_{22} x^{2}y^{2} 
\nonumber \\
m &=& 1 
\nonumber \\
v_{2} &=& 0.5
\nonumber \\
v_{22} &=& 0.05 ~ .
\eea
We work in imaginary time and the convention $\hbar=k_{B}=1$ is used.
For the corresponding quantum action, we make the following ansatz,
compatible with time-reversal symmetry, parity conservation and symmetry under exchange $x \leftrightarrow y$,
\begin{eqnarray}
\tilde{S} &=& \int_{0}^{T} dt ~ 
\frac{1}{2} \tilde{m} (\dot{x}^{2} + \dot{y}^{2}) 
+ \tilde{V}(x,y), ~~~
\nonumber \\
\tilde{V} &=& \tilde{v}_{0} 
+ \tilde{v}_{2} (x^{2} + y^{2}) 
+ \tilde{v}_{22} x^{2}y^{2} 
+ \tilde{v}_{4} (x^{4} + y^{4}) ~ .
\end{eqnarray}
We have determined numerically the parameters of the quantum action 
for transition times from $T=0$ up to $T=4$, where a regime of asymptotic stability is reached. For $T=4$ corresponding to temperature $\tau=0.25$, 
we found 
\begin{eqnarray}
\tilde{m} &=& 0.99814(6)
\nonumber \\  
\tilde{v}_{0} &=& 1.29373(2)
\nonumber \\ 
\tilde{v}_{2} &=& 0.5154(3)
\nonumber \\ 
\tilde{v}_{22} &=& 0.0497(1)
\nonumber \\ 
\tilde{v}_{4} &=& -.0008(7) ~ .
\end{eqnarray} 
We have included terms $\dot{x} \dot{y}$, $xy$, $xy^{3}+x^{3}y$.
Their coefficients were found to be very small (compared to machine precision). 
Also we included terms $x^{2}y^{4} + x^{4}y^{2}$, $x^{4}y^{4}$. Those coefficients were found compatible with zero within error bars.
The quantum action slightly modifies the parameters $\tilde{v}_{2}$ and the parameter $\tilde{v}_{22}$.

Like it was done in sect.[\ref{sec:AnalyticForm}] for the 1-D case, we would like to check also in the 2-D system how well the quantum action reproduces the ground state properties in the zero-temperature limit. 
The classical action considered now is neither rotationally symmetric, nor 
integrable. Nevertheless, we can express the ground state energy and wave function in terms of the quantum action, and compare it with the solution of the Schr\"{o}dinger equation. 
We use the property of $\tilde{V}(\vec{x})$ having a unique minimum at $\vec{x}=0$. In the limit $T \to \infty$ using conservation of energy, $ - \tilde{T}_{kin} + \tilde{V} = \epsilon = \tilde{v}_{0}$, the quantum action can be evaluated along its classical trajectory, with boundary conditions 
$\tilde{\vec{x}}_{cl}(t=0) = \vec{x}_{in}$ and
$\tilde{\vec{x}}_{cl}(t=T) = \vec{x}_{fi}$ to yield
\bea
\label{ActFeynKac}
&& \tilde{\Sigma} \equiv \tilde{S}[\tilde{\vec{x}}_{cl}]|_{\vec{x}_{in},0}^{\vec{x}_{fi},T} 
= \int_{0}^{T} dt ~ \tilde{T}_{kin} + \tilde{V} 
= \int_{0}^{T} dt ~ 2 \tilde{T}_{kin} + \epsilon
\nonumber \\
&& \longrightarrow_{T \to \infty} \tilde{v}_{0} T + \tilde{m} \int_{0}^{T} dt ~ 
(\frac{d \tilde{\vec{x}}_{cl}}{dt})^{2} 
= \tilde{v}_{0} T + \tilde{m} \left( \int_{0}^{T/2} 
+ \int_{T/2}^{T} dt ~ (\frac{d \tilde{\vec{x}}_{cl}}{dt})^{2} \right)
\nonumber \\
&& = \tilde{v}_{0} T + \tilde{m} \left( 
\int_{\vec{x}_{in}}^{0} + \int_{0}^{\vec{x}_{fi}} d \tilde{\vec{x}}_{cl} \cdot 
(\frac{d \tilde{\vec{x}}_{cl}}{dt}) \right) ~ .
\eea
Using the conjecture, Eq.(\ref{DefQuantumAction}), gives the transition amplitude
\be
\label{TransAmplQuantPot}
G(\vec{x}_{fi},T;\vec{x}_{in},0) \longrightarrow_{T \to \infty}
\tilde{Z}_{0} ~ e^{-\tilde{v}_{0} T/\hbar} ~ 
\exp[ - \tilde{m} \int_{0}^{\vec{x}_{fi}} d \tilde{\vec{x}}_{cl} \cdot 
\frac{d \tilde{\vec{x}}_{cl}}{dt} ] ~  
\exp[ - \tilde{m} \int_{\vec{x}_{in}}^{0} d \tilde{\vec{x}} \cdot 
\frac{d \tilde{\vec{x}}_{cl}}{dt} ] ~ , 
\ee
where we subsumed $\tilde{Z}/\tilde{Z}_{0}$ into $\tilde{\Sigma}$ and $\tilde{Z}_{0}$ is a constant of dimension $1/L^{D}$. 
Comparison with the Feynman-Kac formula yields
the ground state energy and wave function,
\be
E_{gr} = \tilde{v}_{0} , ~~~ \psi_{gr}(\vec{x}) = \frac{1}{N} ~ 
\exp[ - \tilde{m} \int_{0}^{\vec{x}} d \tilde{\vec{x}}_{cl} \cdot 
\frac{d \tilde{\vec{x}}_{cl}}{dt} ] ~ ,
\ee
where $\tilde{\vec{x}}_{cl}$ is the trajectory from 
$\tilde{\vec{x}}_{cl}(t=T/2) = o$ to
$\tilde{\vec{x}}_{cl}(t=T) = \vec{x}$. 
The ground state wave function thus obtained from the 
quantum action has been compared to the solution of the Schr\"{o}dinger equation. This is shown in Fig.[2]. 
We also compared the ground state energy. The solution of the stationary Schr\"odinger equation yields $E_{gr}=1.01207$. Like in the 1-D system 
(see Eq.(\ref{GroundStateFeynKac}))  
the ground state energy is given by the minimum of the quantum potential.
Extrapolation of $\tilde{v}_{0}$ to large $T$ yields
$\tilde{v}_{0} \to 1.0127$.
One observes that the quantum action describes well the physics at zero temperature also of the 2-D anharmonic oscillator. Compared to the 1-D system, 
however, the error in the wave function is larger by one order of magnitude. 
A reason is certainly due to the fact that in 1-D we have an analytical expression (\ref{GroundStateFeynKac}) for the wave function from the quantum action, while in 2-D this is not available.

Now let us look at the chaotic behavior of the classical and the quantum 
action. We have computed Poincar\'e sections from the classical action 
and from the quantum action for a variety of temperatures. 
The equations of motion of the quantum action have been solved using a 4-th order Runge-Kutta algorithm, and Henon's algoritm was used to compute the Poincar\'e sections. 
Let us discuss what we expect.
First, the classical system under consideration is 
chaotic due to the anharmonic 
interaction term. This term is small compared to the harmonic 
oscillator term. This property proliferates to the quantum potential.
Here "smallness" is meant in the following sense: The quantum system at 
zero temperature is determined by the ground state properties. 
We have computed 
$r_{0}=\langle r \rangle_{gr}$, the radius of the ground state wave 
function (corresponding to the Bohr radius $a_{0}$ in the hydrogen atom).
We obtain $r_{0} = 0.8766$. 
It turns out that the anharmonic term is smaller than the harmonic term 
for any length $r < r_{0}$.
Thus one expects roughly the same "amount" of chaotic behavior in the 
quantum system as in the classical system. 
Second, renormalisation (i.e. change of action parameters when going from 
the classical action to the quantum action) tends to increase harmonic terms, 
i.e. drive the system to a Gaussian fixed point. At a Gaussian fixed point the system is integrable. i.e. non-chaotic.
This effect has been observed in a number of numerical studies in 1-D systems \cite{Jirari:01a,Jirari:01c}. In a system with a symmetric double well 
potential this effect was found to lead to a softening of the double well shape of the quantum potential (lower barrier, more narrow minima, 
even vanishing of the double well). 
This translates to a softening or even an "evaporation" of quantum instantons 
compared to the classical instantons \cite{Jirari:01c}. 
This effect is present also in the system under consideration here. We find
$\delta \tilde{v}_{2} = \tilde{v}_{2} - v_{2} = 0.0154(3)$ compared to 
$\delta \tilde{v}_{22} = \tilde{v}_{22} - v_{22} = -.0003(1)$. 
We expect that this effect may lead to a slight softening of the chaotic 
behavior in the quantum system.
Third, a weak anharmonic term means that the quantum potential reaches 
a regime of asymptiotic stability as function of $T$ already for small $T$ 
($T \approx 4$). That is, when going from high to low temperature, 
already at relatively high temperature the asymptotic regime (Feynman-Kac limit) is reached.

We computed the Poincar\'{e} sections for the quantum action at temperature $\tau=0.25$ (corresponding to $T=4$) for energies $E=10, 20, 50$. 
One should keep in mind that the classical action at $T=0$ is equivalent to a quantum action at temperature $\tau=\infty$.
Those Poincar\'e sections compared to their counter part from the classical action are shown in Figs.[3-5]. One observes that the quantum system also displays chaos, and the Poincar\'{e} sections are slightly different from those of the classical action. Like in the classical case also in the quantum system the "amount" of chaoticity increases with increasing energy.   
Moreover, the difference between classical and quantum Poincar\'e sections becomes more accentuated with increase of energy.
For comparison, we also display in Fig.[6] the Poincar\'e sections of the quantum action at $E=50$ for temperature $\tau=2$ and $\tau=0.5$, i.e. 
closer to the high temperature limit.

\section{Summary}
\noindent We have discussed the use of the quantum action, which can be 
considered as a renormalized classical action at finite temperature. Based on the quantum action we suggest a proper definition 
of quantum chaos and explore it numerically.
As example, we have considered in 2-D the harmonic oscillator with a weak anharmonic coupling and computed the quantum action at 
various temperatures. We compared Poincar\'{e} sections 
of the quantum action with those of the classical action and found some difference in their chaotic behavior. \\

\noindent {\bf Acknowledgements} \\ 
H.K. and K.M. are grateful for support by NSERC Canada. 
X.Q.L. has been supported by NSF for Distinguished Young Scientists of China, by Guangdong Provincial NSF and by the Ministry of Education of China.

\newpage
\begin{flushleft}
{\bf Figure Caption}
\end{flushleft}
\begin{description}
\item[{Fig.1}]
1-D anharmonic oscillator. Ground state wave function from Schr\"odinger equation (open dots) and from quantum action (squares). Their difference is shown in the lower part. 
\item[{Fig.2}]
Ground state wave function of 2-D anharmonic oscillator. $x$ indicates the location of cuts through the wave function. The upper part shows the solution from Schr\"odinger equation (full line) and from quantum action (different symbols). The difference is shown in the lower part.
\item[{Fig.3}]
2-D anharmonic oscillator. Poincar\'e sections at energy $E=10$. 
Upper part classical action, lower part quantum action at temperature $\tau=0.25$. 
\item[{Fig.4}]
Same as Fig.[3], but $E=20$. 
\item[{Fig.5}]
Same as Fig.[3], but $E=50$.
\item[{Fig.6}]
Poincar\'e sections of quantum action at $E=50$, $\tau=2$ and $\tau=0.5$. 

\end{description}

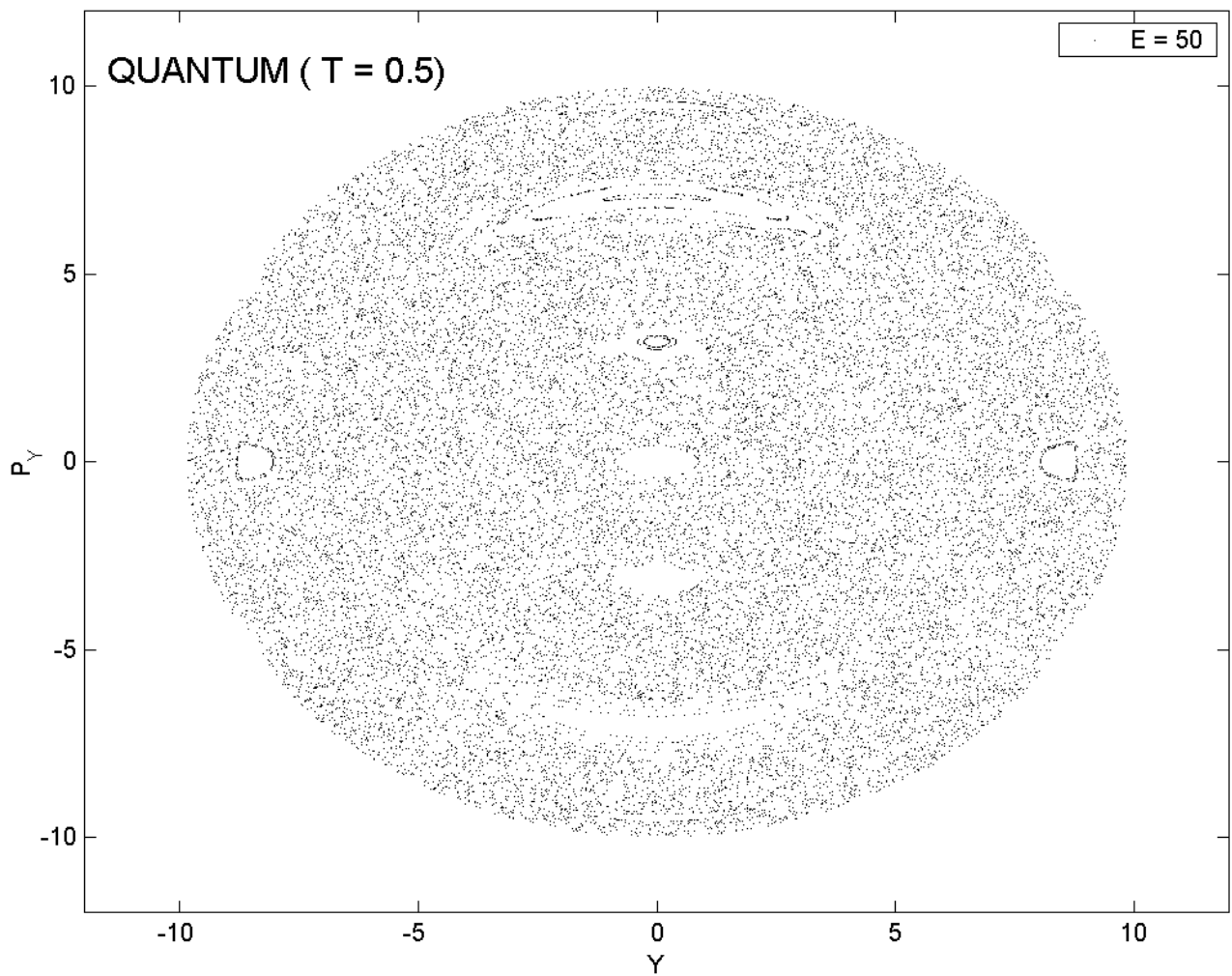
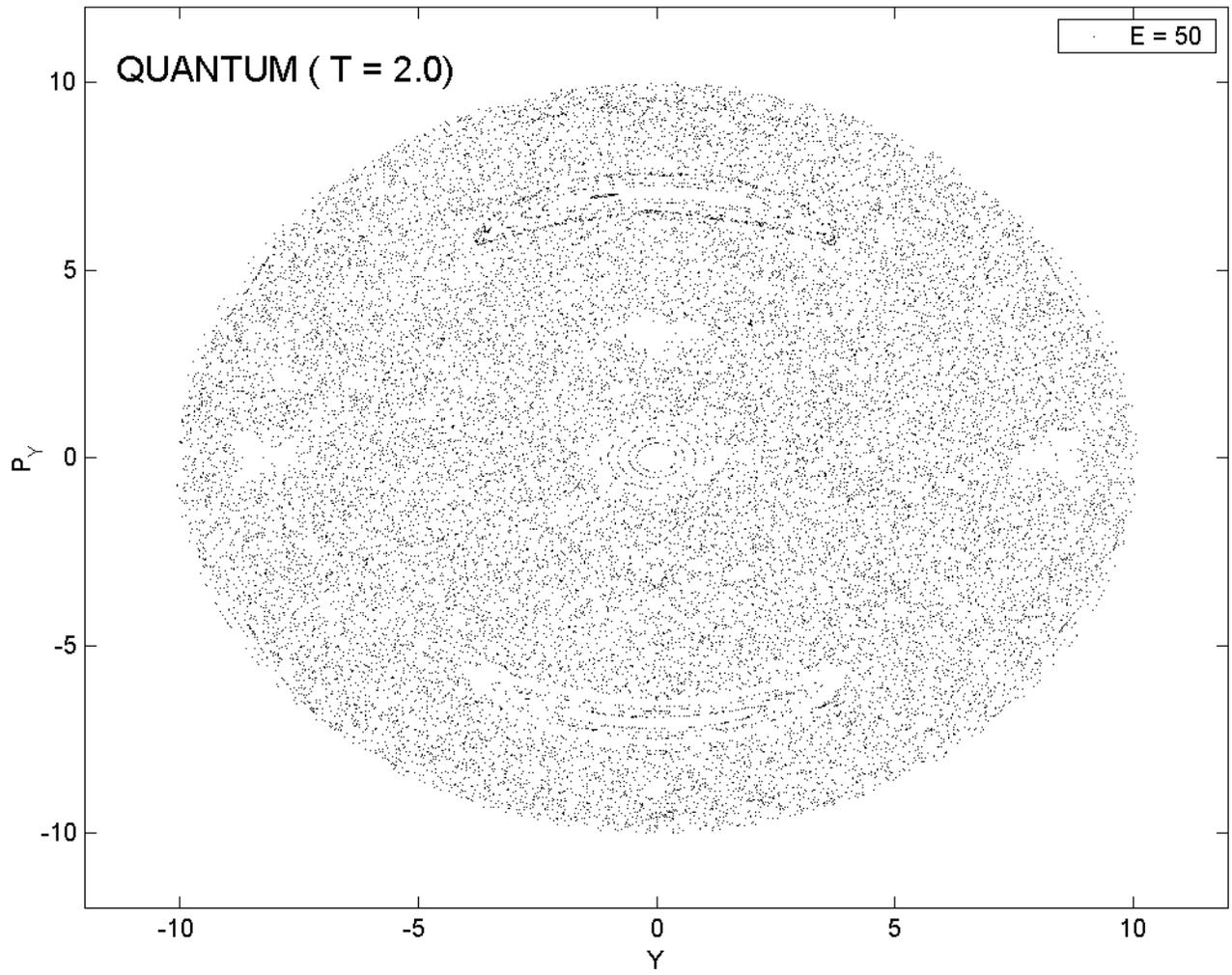
\end{document}